\documentclass[11pt,a4paper]{scrartcl}

\usepackage{ILD}

\usepackage[symbol]{footmisc}
\usepackage{feynmf}
\usepackage{textpos}
\usepackage{lineno}
\usepackage[sorting=none]{biblatex}

\usepackage{physics}
\usepackage{caption}
\usepackage{subcaption}
\usepackage{csquotes}
\title{Prospects of fast timing detectors for particle
identification at future Higgs factories}

\ildproc{phys}{2021}{006}
\date{\today}
\addauthor{Bohdan Dudar}{\institute{1} \institute{2}}
\addauthor{Jenny List}{\institute{1}}
\addauthor{Ulrich Einhaus}{\institute{1} \institute{2}}
\addauthor{Remi Ete}{\institute{1}}

\addinstitute{1}{DESY}
\addinstitute{2}{Hamburg University}

\abstract{
We present an overview of a study on precise mass reconstruction and identification of charged hadrons ($\pi^{\pm}$, $K^{\pm}$, $p$) using time-of-flight measurements in the electromagnetic calorimeter of a typical Higgs factory detector. 
Time-of-flight measurements can take advantage of fast timing Si sensors with a time resolution in the order of 10\,ps. A precise time-of-flight measurement might contribute to the kaon mass determination and can improve particle identification in the momentum regions inaccessible for the $dE/dx$ method. In this contribution, we discuss the current status and the challenges of the time-of-flight approach for a precise reconstruction of charged hadron masses.
}

\titlecomment{This work was carried out in the framework of the ILD concept group. Talk presented at the International Workshop on Future Linear Colliders (LCWS2021), 15-18 March 2021. C21-03-15.1.}

\graphicspath{ {./logos/}{./figures/} }

\begin{document}

% generates the title page
\titlepage

% include source for sections
\section{Introduction}

The performance requirements for detectors at a future Higgs factory have been known for a long time in terms of track momentum resolution, jet energy resolution, impact parameter resolution and hermeticity. More recently it has been realised that the particle identification (PID) ability to distinguish different kinds of charged hadrons provides important additional information~\cite{tof_ref1, tof_ref2,tof_ref3}. Some proposed detector concepts like ILD~\cite{TDR, IDR} at the ILC~\cite{ilc} or the CEPC~\cite{cepc2} detector offer PID via the specific energy loss ($dE/dx$) in their gaseous main tracking devices, which could be complemented by time-of-flight (TOF) measurements. For the other detector concepts which rely on silicon tracking only, TOF would be the only possibility for charged hadron PID. TOF measurements could be implemented e.g. with fast timing silicon sensors placed in the innermost layers of the electromagnetic calorimeter (ECal) or in the outermost tracker layer.

We present here a study based on the ECal, assuming single hit time resolutions between 10 and 100\,ps, as they can be reached by modern silicon sensor technologies~\cite{LGAD}. With this kind of resolutions, TOF-based PID would allow to identify charged hadrons with momenta up to about 5 GeV, exactly in the region where the Bethe-Bloch bands overlap, prohibiting the identification via $dE/dx$, see Fig.\,8.6 in~\cite{IDR}. The TOF-based PID approach is relevant for any Higgs factory, however, for this study we use the ILD concept and its detailed and well-established full simulation as a showcase.

In addition to the PID, TOF could potentially also be employed to improve our knowledge about the kaon mass. The two most precise measurements of the kaon mass, both from spectroscopy of kaonic atoms, are discrepant and the PDG quotes their average as $493.677 \pm 0.013 \mathrm{MeV}$~\cite{pdg}. If a mass measurement from TOF could reach a precision of about 10 keV, future Higgs factories could contribute to clarifying this discrepancy.

The results presented here were obtained using an $e^+e^- \rightarrow Z \rightarrow q\bar{q}$ sample from the IDR MC production of ILD~\cite{Production_2018} ($\sqrt{s}=500$\,GeV, iLCSoft v02-00-02, ILD\_l5\_o1\_v02). Details of this production and the employed standard reconstruction can be found in~\cite{IDR}, and we only point out that the clustering of the calorimeter hits and the matching of tracks and clusters to particle flow objects (PFOs) were performed with the Pandora particle flow algorithm~\cite{Pandora}. As a simple test ground for our methods, only PFOs with exactly on track and one cluster in the barrel part of the ECal were considered. Furthermore, the tracks were required to have not fewer than 200 of the maximum possible 220 TPC hits and have track parameters at the interaction point (IP) $\abs{d_{0}} < 10$\,mm, $\abs{z_{0}} < 20$\,mm to ensure the quality of the tracks. For the studies with photons we required a cluster longitudinal position of $\abs{z_{\mathrm{cluster}}} < 2200$\,mm and a true vertex position within 500\,$\mu$m to the IP.
\section{The basic principle of the TOF particle ID}

Measurements of the momentum $p$ and the velocity $\beta$ in natural units of a charged particle determine its mass via the relativistic momentum formula depicted in Equation~\ref{eq:1}:

\begin{equation}
    \label{eq:1}
    m = \frac{p}{\beta}\sqrt{1 -\beta^2}  
\end{equation}

The momentum is measured in the tracking system. The track reconstruction of ILD is described in~\cite{track}. It is based on a Kalman filter which provides a helix-based parametrisation of the particle's trajectory at every hit, as well as at the IP and the calorimeter front face. Of particular importance for the momentum reconstruction are the curvature $\Omega$ and the dip angle $\lambda$ of the helix. With these and the strength of the solenoidal magnetic field $B_z$, the momentum is determined via Equation~\ref{eq:2}:

\begin{equation}
    \label{eq:2}
    p = e \abs{\frac{B_z}{\Omega}}\sqrt{1 + \tan^2\lambda}
\end{equation}

The velocity $\beta$ of the particle is calculated from the ratio of the track length $\ell_{\mathrm{track}}$ to the time-of-flight $\tau$:

\begin{equation}
    \label{eq:3}
    \beta = \frac{\upsilon}{c} = \frac{\ell_{\mathrm{track}}}{\tau \cdot c}
\end{equation}

\begin{figure}[!htbp]
    \centering
    \includegraphics[width=0.75\textwidth]{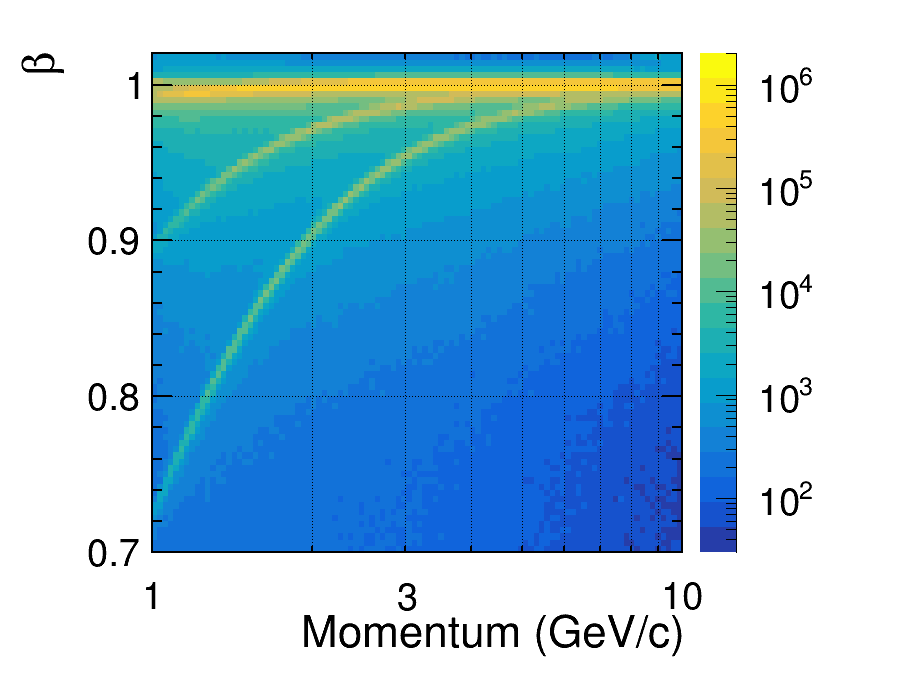}
    \caption{$\beta$-versus-momentum plane. The three separate bands correspond to $\pi^{\pm}$, $K^{\pm}$ and $p$ particles. The bands are easily separable up to momenta of 3-5\,GeV. The plot is made with $\Omega_{\mathrm{IP}}$, $\lambda_{\mathrm{IP}}$, $\tau_{\mathrm{avg}}$} and assuming perfect time resolution. See Section\,2 for the details.
    \label{fig:beta_bands}
\end{figure}

Particles are identified by their corresponding $p$ and $\beta$ which layout in separate bands that correspond to the different types of hadrons~\cite{uli_talk}. An example of these bands corresponding to $\pi^{\pm}$, $K^{\pm}$ and $p$ is shown in Figure~\ref{fig:beta_bands}. The time-of-flight is estimated based on the timing information from the ECal hits. We will discuss the procedures involved in the section below in more detail.

The track parameters also serve to estimate the track length $\ell_{\mathrm{track}}$ with Equation~\ref{eq:4}. Therein, $\varphi_{\mathrm{ECal}}$ and $\varphi_{\mathrm{IP}}$ are the angles of the helix direction at the entry point to the ECal and the point of the closest approach to the IP, respectively.

\begin{equation}
    \label{eq:4}
    \ell_{\mathrm{track}} = \abs{\frac{\varphi_{\mathrm{ECal}} - \varphi_{\mathrm{IP}}}{\Omega}} \sqrt{1 + \tan^2\lambda}
\end{equation}

\subsection{The choice of the track parameters}

In the formulae above, the track parameters $\Omega$ and $\lambda$ that are used for the momentum and track length calculation are assumed to be constant. However, due to energy loss they change as the track propagates through the tracking system. In Section\,4, we will compare the final mass spectrum obtained via Equation~\ref{eq:1} when calculating $\ell_{\mathrm{track}}$ and $p$ from the track parameters at the IP ($\Omega_{\mathrm{IP}}$, $\lambda_{\mathrm{IP}}$) and at the entry point to the calorimeter ($\Omega_{\mathrm{calo}}$, $\lambda_{\mathrm{calo}}$). In principle, the track length can be calculated with better precision directly from the Kalman filter that we use for track reconstruction. We plan to study this approach in the future. Also, vertex information is not yet taken into account in this study and it is always assumed that tracks start at the point of the closest approach to the IP.

\subsection{The choice of the TOF estimator}

We test four methods to estimate the particle-level TOF from the time information given by the ECal hits. The ECal hit time is considered to be the time of the earliest MC contribution to the energy deposition in the hit. To simulate a finite hit time resolution we apply Gaussian smearing with a standard deviation equal to the assumed time resolution. For conceptual studies, this smearing is omitted in some cases, labeled as "perfect time resolution". Further digitization effects from the electronics were not taken into account in the simulation and will be addressed in future studies.

The two easiest approaches are to take either the time of the ECal hit closest to the point where the extrapolated track enters the calorimeter ($\tau_{\mathrm{closest}}$) or the time of the earliest hit in the cluster ($\tau_{\mathrm{earliest}}$). To extrapolate the time-of-flight to the ECal surface, instead of the ECal hit position, one needs to correct for the distance between the track's entry point and the center of the ECal hit ($d_{\mathrm{hit,entry}}$). Then, the corrected TOF ($\tau_{\mathrm{corr}}$) for each hit is given by Equation~\ref{eq:5}:

\begin{equation}
    \label{eq:5}
    \tau_{\mathrm{corr}} = t_{\mathrm{hit}} - \frac{d_{\mathrm{hit,entry}}}{c}\mathrm{,}
\end{equation}
where $t_{\mathrm{hit}}$ is the time measured in the ECal hit. This requires an assumption on the speed by which the particle travels inside the calorimeter and/or the shower propagates. For practical purposes, the speed of light is assumed here, which can 
lead to biases.

\begin{figure}[!htbp]
    \centering
    \begin{subfigure}[t]{0.37\textwidth}
        \centering
        \includegraphics[width=\textwidth]{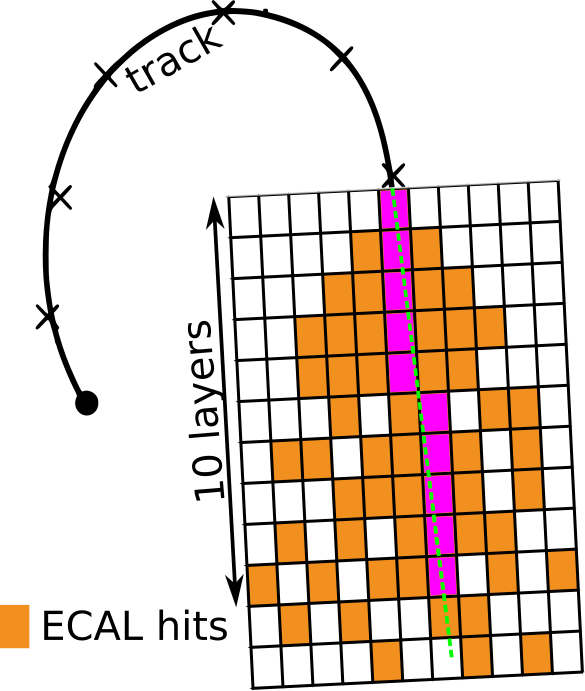}
        \caption{Sketch of the ECal hit selection for the TOF calculation with multi-hit methods.}
        \label{fig:02a}
    \end{subfigure}
    \hfill
    \begin{subfigure}[t]{0.53\textwidth}
        \centering
        \includegraphics[width=\textwidth]{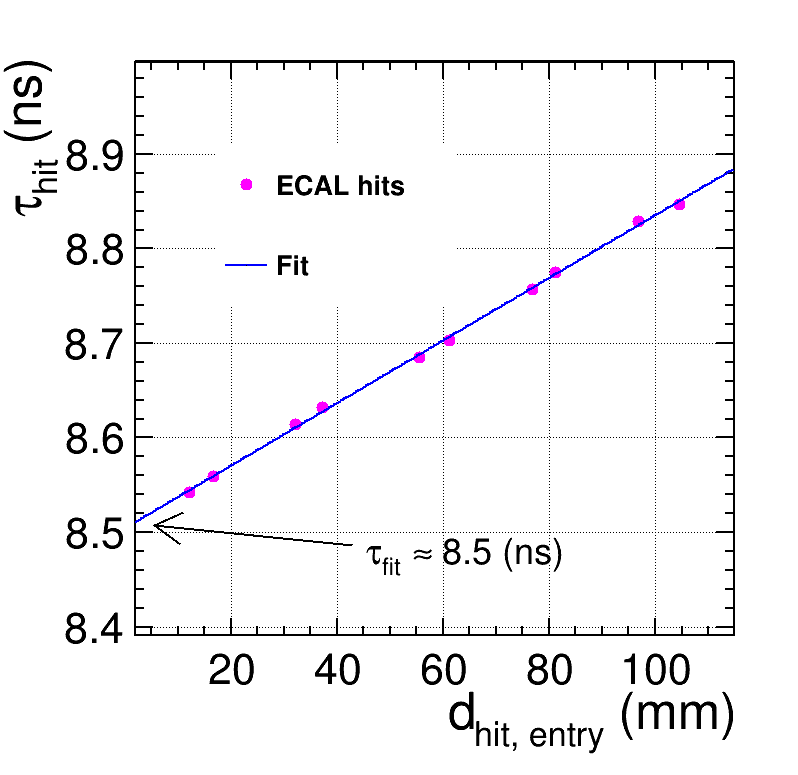}
        \caption{Linear fit of time versus distance to the entry point for the selected ECal hits.}
        \label{fig:02b}
    \end{subfigure}
    \caption{Illustration of multi-hit TOF estimators.}
\end{figure}

As relying only on a single hit suffers from fluctuations in the shower development and the time measurement, we also study TOF estimators that rely on multiple hits in the shower.
The multi-hit estimators combine information from the first 10 layers of the ECal, thereby selecting in each layer the hit which is closest to the extrapolated track, as illustrated in Fig.~\ref{fig:02a}. For charged hadrons, this approximates the MIP part of the cluster, before the hadron actually starts to shower. In the future, this could be replaced by a dedicated shower start finder. The time information of the selected hits can then be combined either by averaging the corrected times at the calorimeter entrance ($\tau_{\mathrm{avg}}$), or by fitting the velocity of the particle's propagation ($\tau_{\mathrm{fit}}$).

In case of the fit, a linear function is used to describe the time of the hits as a function of the distance to the entry point. An example of such a fit is shown in Fig.~\ref{fig:02b}.

\section{Results}

In this section we present results of the study of different track parameter options and TOF estimators in terms of precise mass reconstruction for charged hadrons.

\subsection{Mass reconstruction}

\begin{figure}[!htbp]
    \centering
    \begin{subfigure}[t]{0.45\textwidth}
        \centering
        \includegraphics[width=\textwidth]{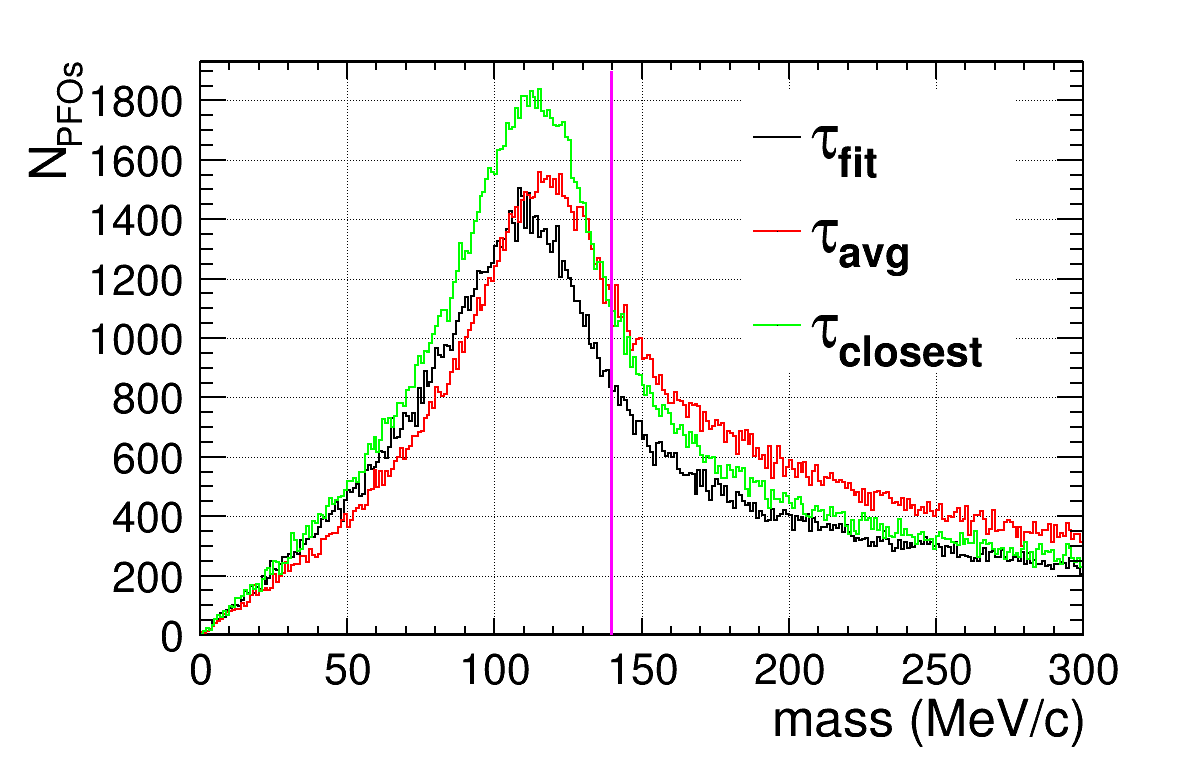}
        \caption{$\Omega_{\mathrm{IP}}$, $\lambda_{\mathrm{IP}}$ track parameters for $\pi^{\pm}$.}
        \label{fig:03a}
    \end{subfigure}
    \hfill
    \begin{subfigure}[t]{0.45\textwidth}
        \centering
        \includegraphics[width=\textwidth]{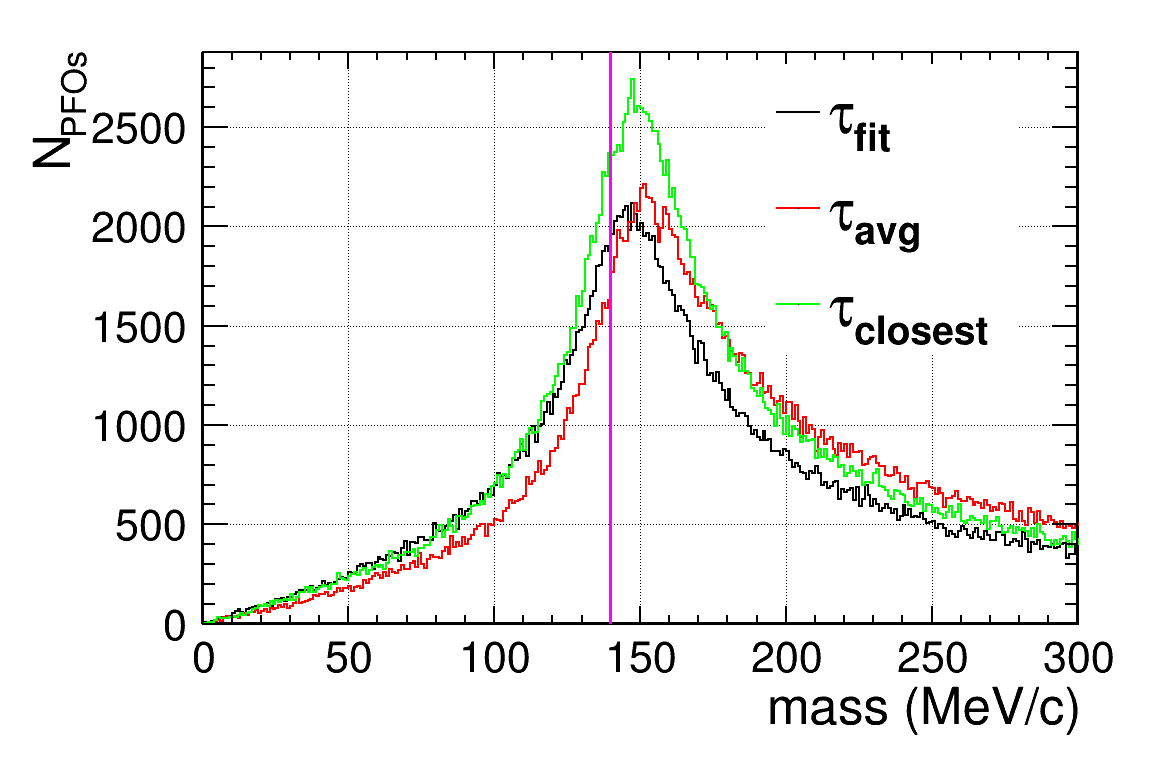}
        \caption{$\Omega_{\mathrm{calo}}$, $\lambda_{\mathrm{calo}}$ track parameters for $\pi^{\pm}$.}
        \label{fig:03b}
    \end{subfigure}
    %\bigskip
    \begin{subfigure}[t]{0.45\textwidth}
        \centering
        \includegraphics[width=\textwidth]{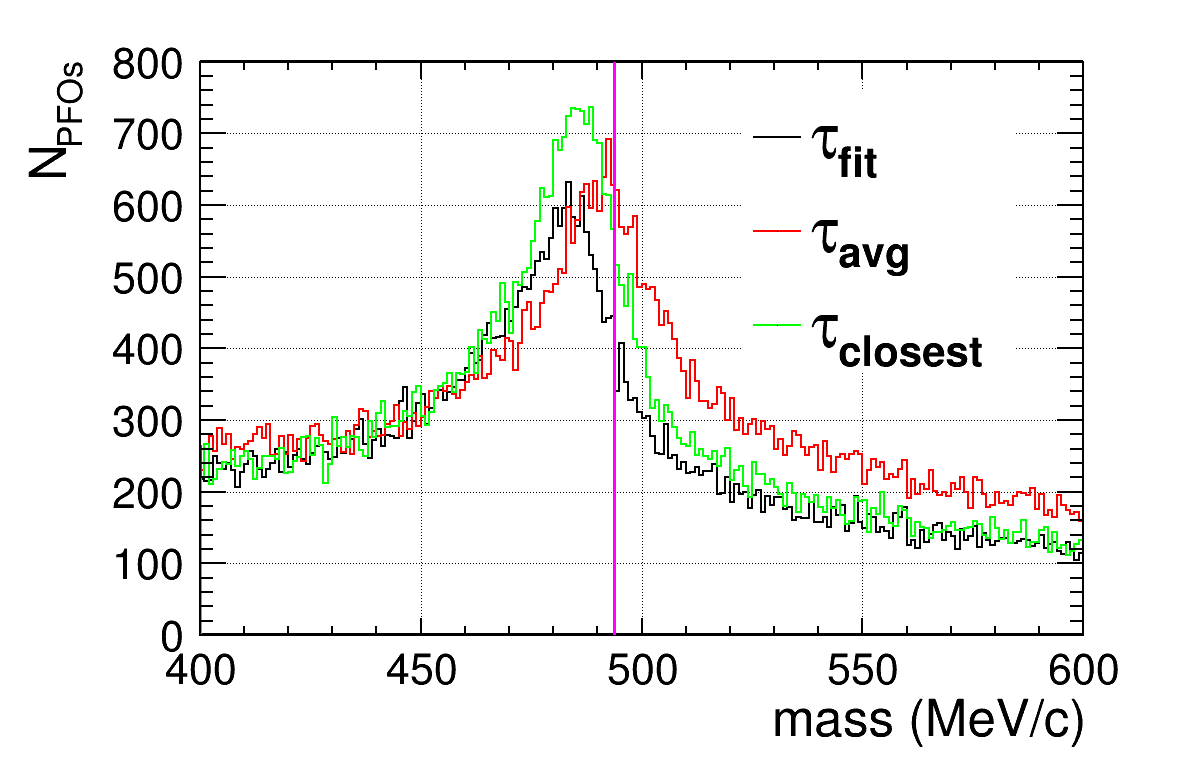}
        \caption{$\Omega_{\mathrm{IP}}$, $\lambda_{\mathrm{IP}}$ track parameters for $K^{\pm}$.}
        \label{fig:03c}
    \end{subfigure}
    \hfill
    \begin{subfigure}[t]{0.45\textwidth}
        \centering
        \includegraphics[width=\textwidth]{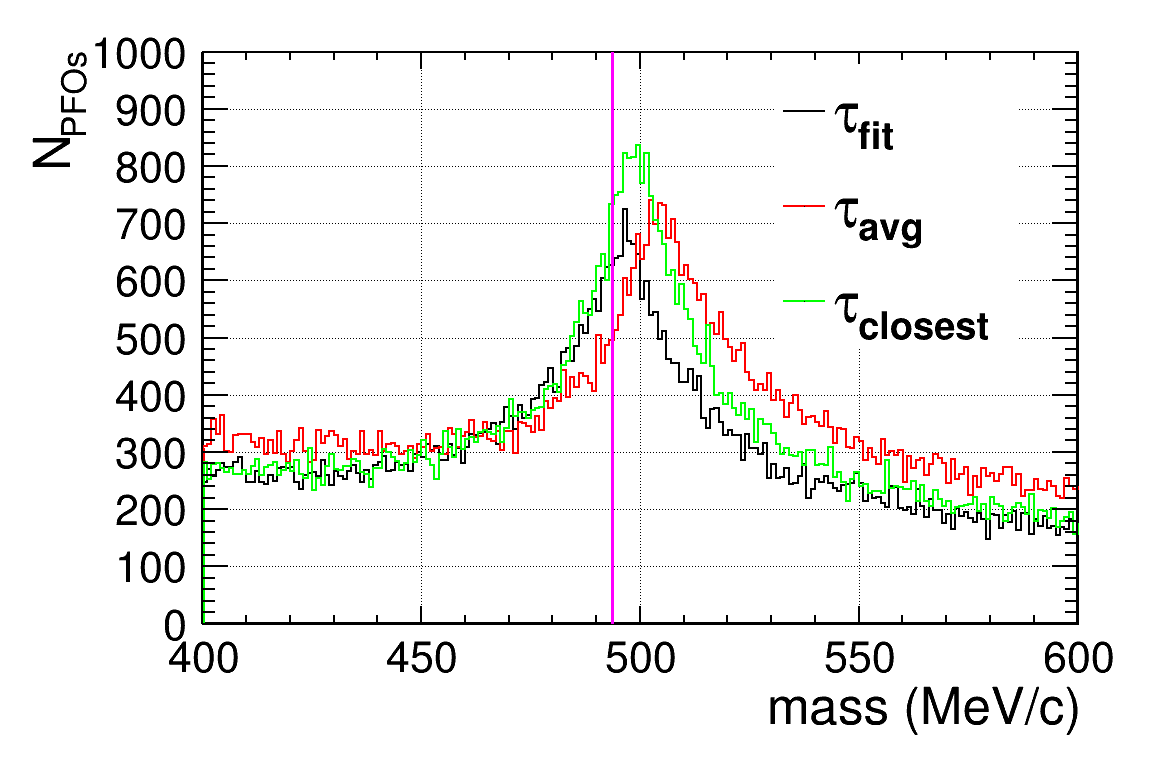}
        \caption{$\Omega_{\mathrm{calo}}$, $\lambda_{\mathrm{calo}}$ track parameters for $K^{\pm}$.}
        \label{fig:03d}
    \end{subfigure}
    %\bigskip
    \begin{subfigure}[t]{0.45\textwidth}
        \centering
        \includegraphics[width=\textwidth]{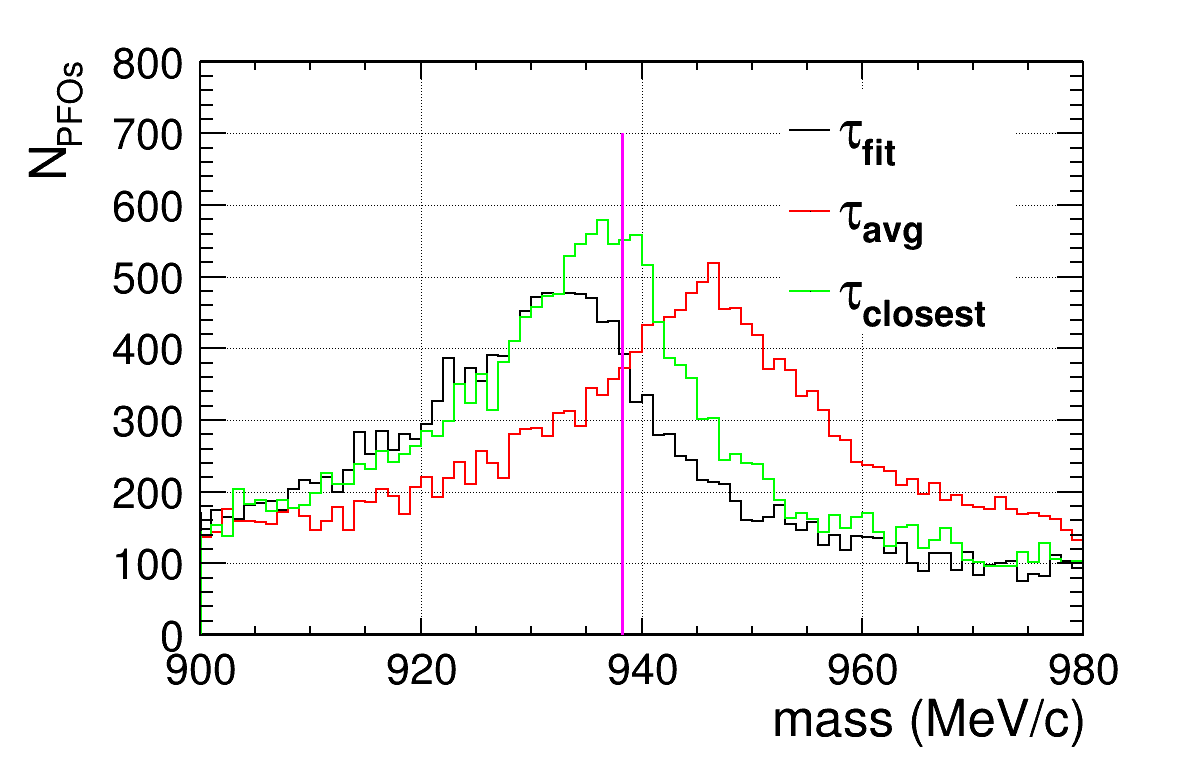}
        \caption{$\Omega_{\mathrm{IP}}$, $\lambda_{\mathrm{IP}}$ track parameters for $p$.}
        \label{fig:03e}
    \end{subfigure}
    \hfill
    \begin{subfigure}[t]{0.45\textwidth}
        \centering
        \includegraphics[width=\textwidth]{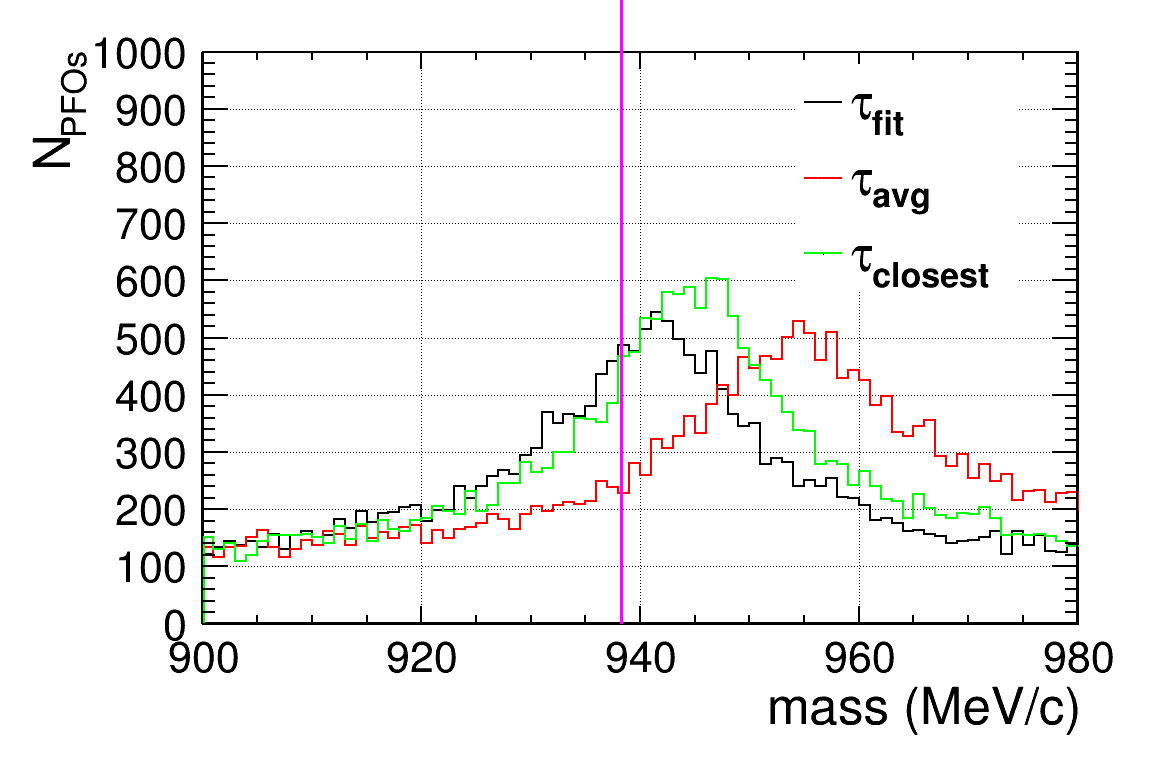}
        \caption{$\Omega_{\mathrm{calo}}$, $\lambda_{\mathrm{calo}}$ track parameters for $p$.}
        \label{fig:03f}
    \end{subfigure}
    \caption{Mass spectrum for $\pi^{\pm}$, $K^{\pm}$, $p$ using different track parameters and TOF estimators for the mass calculation.}
    \label{fig:03}
\end{figure}

\begin{figure}[!htbp]
    \centering

    \includegraphics[width=0.75\textwidth]{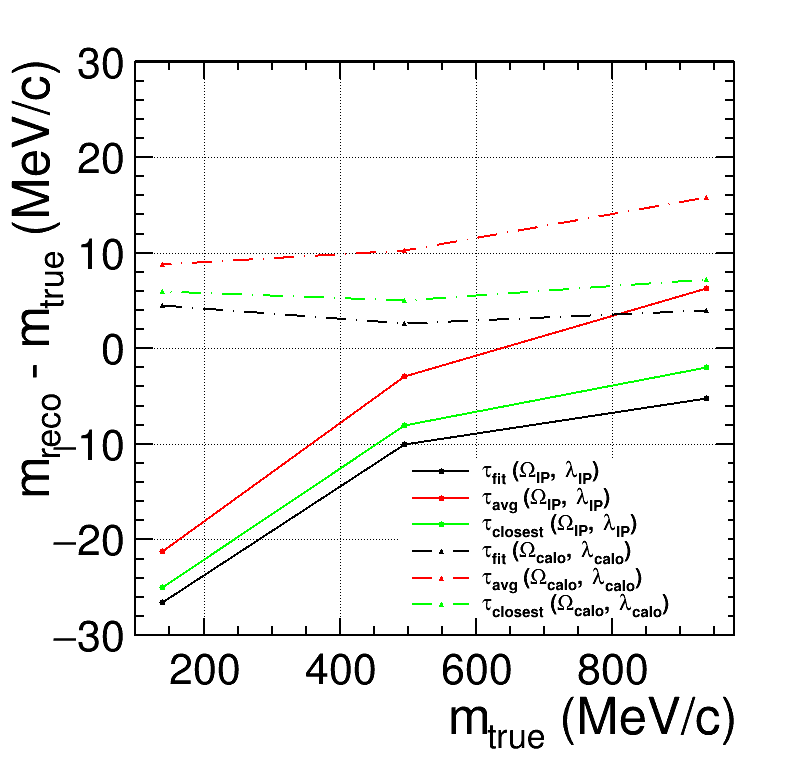}

    \caption{The bias in the reconstructed mass by different methods that combine hit-level time information to a cluster-level time-of-flight. $\tau_{\mathrm{fit}}$ estimator with track parameters at the calorimeter surface shows the smallest bias which is also least deviant between particles.}
    \label{fig:04}
\end{figure}

Reconstructed mass peaks of the $\pi^{\pm}$, $K^{\pm}$, $p$ particles are shown in Fig.~\ref{fig:03} using different track parameters and TOF estimators evaluated on true hit times, without smearing for the time resolution.
Different methods show different peak positions of the mass distribution which differ from the PDG values at the level of $O$(10\,MeV).
We define the mass bias as the difference between the peak position and the PDG value. The peak positions are determined with a Gaussian fit in a local region around the observed maximum. The mass biases for all particles and all methods are combined in one summarizing plot in Fig.~\ref{fig:04}. 
Using the track parameters at the calorimeter surface leads to a consistent bias for all particles, while taking the track parameters at the IP leads to a lower reconstructed mass for $\pi^{\pm}$ and a higher reconstructed mass for $p$ than expected.
In the case of using track parameters at the calorimeter surface, $\tau_{avg}$ shows the largest bias for all particles. The explanation for this seems to be the assumption of the speed-of-light propagation of the shower, which results in a larger bias compared to the other estimators.
The estimator $\tau_{\mathrm{fit}}$ in combination with the track parameters at the calorimeter impact point shows the smallest bias, but still a remaining bias of 3-4\,MeV which is two orders of magnitude larger than the precision we want to achieve $O$(10\,keV).

For further improvement one needs to find a deeper understanding of the cause of the bias by investigating each of the possible major contributors, i.e. momentum, track length and TOF estimation, separately.

\subsection{Study of TOF estimators with photons}
The bias of the reconstructed mass is caused either by the momentum measurement, the track length measurement or the TOF estimation. It is not a trivial task to study them separately with charged hadrons. However, one can use photons to make track length and momentum calculations trivial and consider only TOF effects.

Photons travel in a straight line with no energy loss in the tracking system and at a constant velocity $c$. This allows to calculate the true track length easily and then calculate the true TOF $\tau_{\mathrm{true}}$ with Equation~\ref{eq:6}:

\begin{equation}
    \tau_{\mathrm{true}} = \frac{\ell_{\mathrm{track, true}}}{c}
    \label{eq:6}
\end{equation}

\begin{figure}[!htbp]
    \centering
    \includegraphics[width=0.75\textwidth]{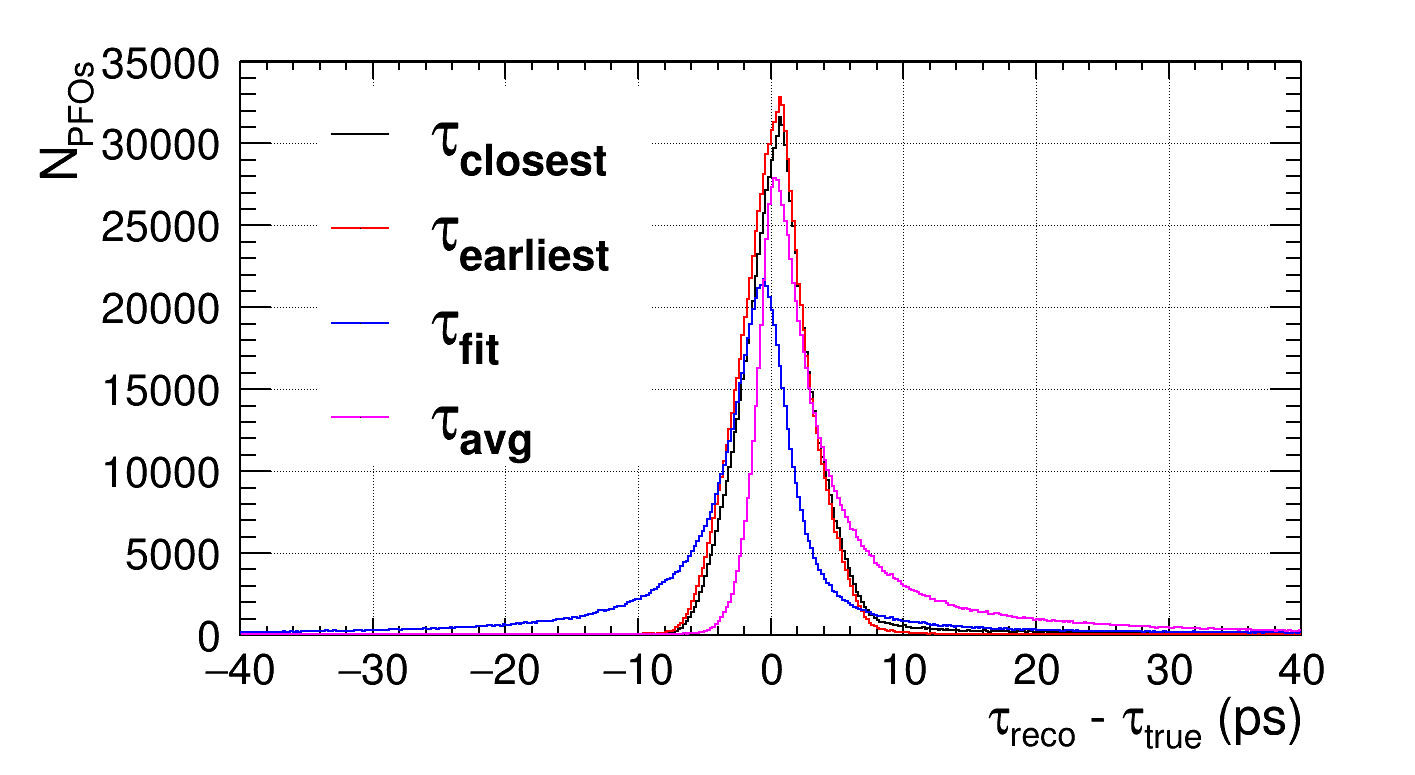}
    \caption{The timing biases of different TOF estimators for photon clusters with perfect time resolution.}
    \label{fig:05}
\end{figure}

\begin{figure}[!htbp]
    \centering
    \includegraphics[width=0.75\textwidth]{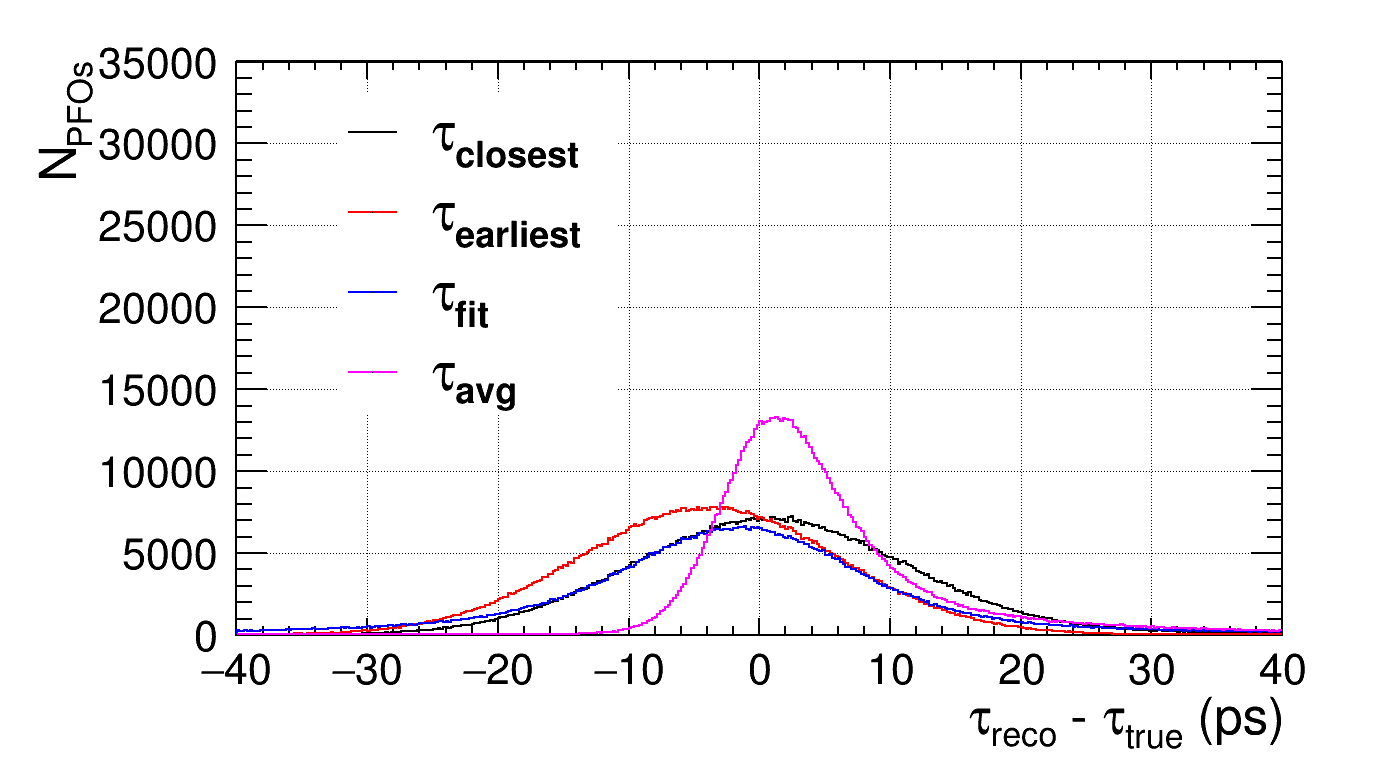}
    \caption{The timing biases of different TOF estimators for photon clusters with 10\,ps time resolution.}
    \label{fig:06}
\end{figure}

The timing bias results for the different TOF estimators are shown in Figs.~\ref{fig:05} and~\ref{fig:06} for perfect and 10\,ps time resolution, respectively.
The least biased and most precise methods are $\tau_{\mathrm{closest}}$ and $\tau_{\mathrm{earliest}}$, respectively, but they degrade very fast when applying a finite time resolution as seen in Fig.~\ref{fig:06}. At 10\,ps time resolution they show a comparable level of precision to $\tau_{\mathrm{fit}}$, while $\tau_{\mathrm{avg}}$ performs significantly better than the others.
From the multi-hit methods, $\tau_{\mathrm{fit}}$ shows the smallest bias, albeit with some sacrifice on the precision compared to the $\tau_{\mathrm{avg}}$, especially for larger time resolutions. In future studies, we want to seek ways to improve the $\tau_{\mathrm{fit}}$ precision with applied hit time resolution.
These results can be used to further study and improve weak points of all TOF estimators or to calibrate them to reduce the bias of the reconstructed mass of charged hadrons.

\begin{figure}[!htbp]
    \centering
    \includegraphics[width=0.75\textwidth]{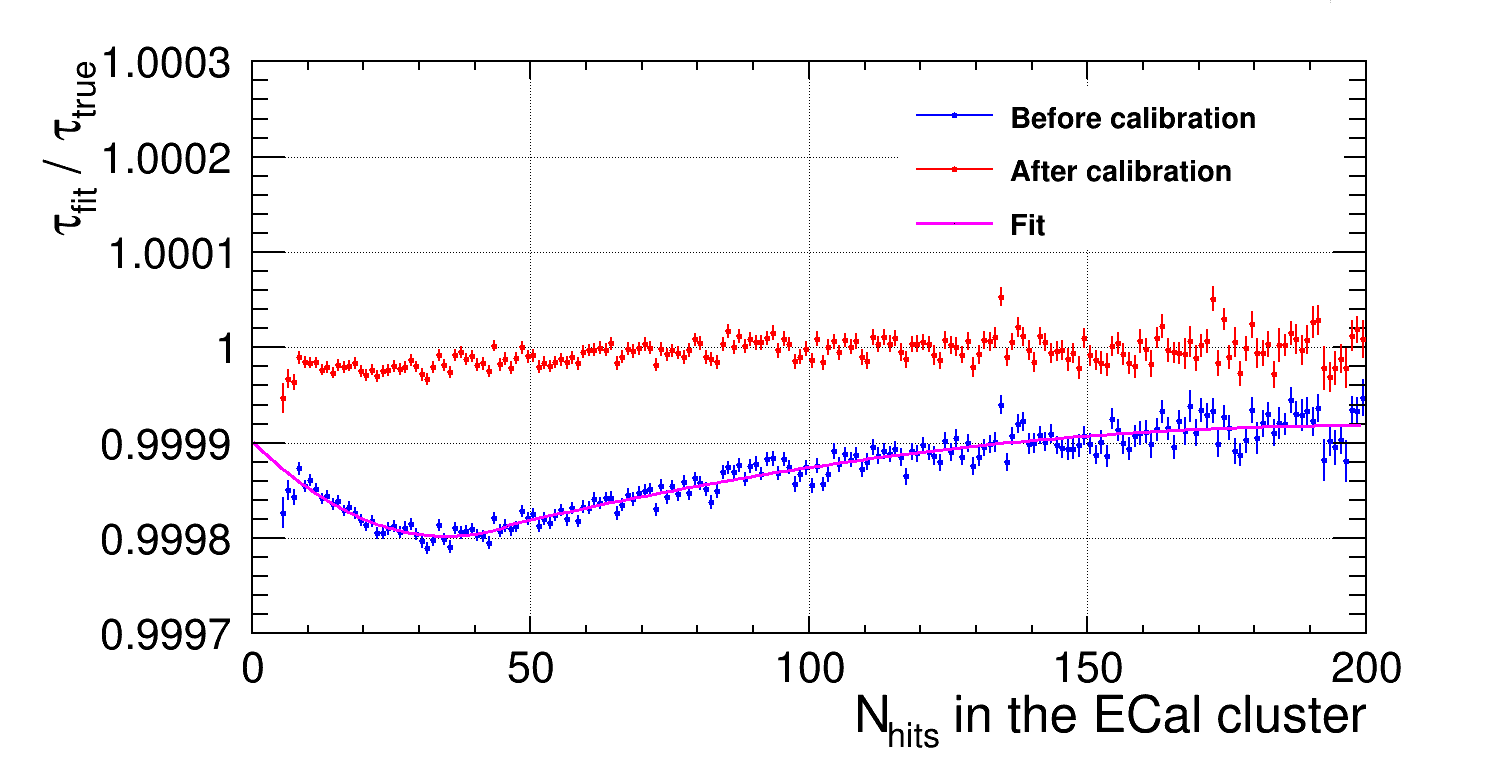}
    \caption{Correlation of the ratio of the reconstructed $\tau_{\mathrm{fit}}$ to the $\tau_{\mathrm{true}}$ with the number of hits in the ECal cluster. The magenta line is used as a calibration curve.}
    \label{fig:07}
\end{figure}

\begin{figure}[!htbp]
    \centering
    \includegraphics[width=0.75\textwidth]{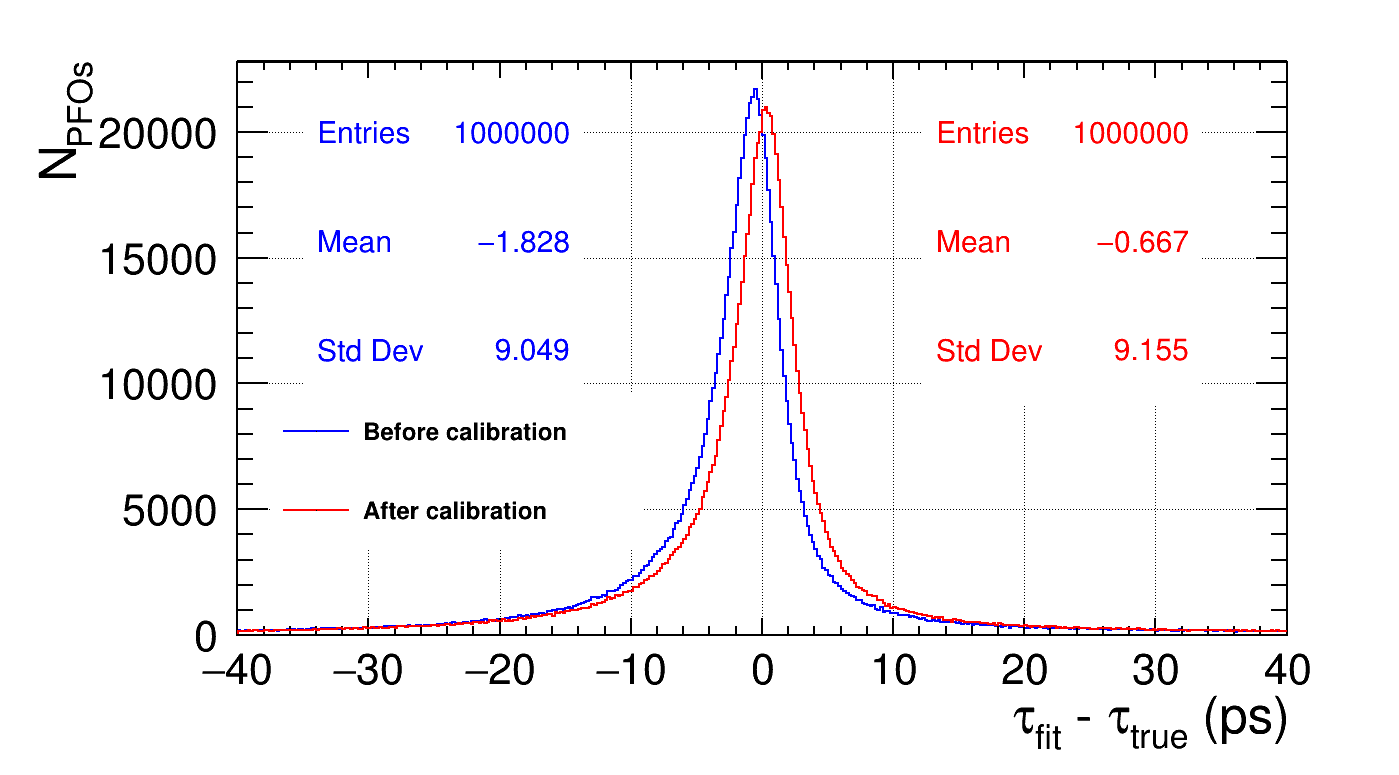}
    \caption{The timing bias distribution of $\tau_{\mathrm{fit}}$ before and after calibration. The mean of the bias is reduced by a factor of 3.}
    \label{fig:08}
\end{figure}

Here we present the example of the calibration procedure and results of the timing bias reduction for the $\tau_{\mathrm{fit}}$ estimator. We used the ratio $\tau_{\mathrm{fit}}/\tau_{\mathrm{true}}$ as a function of the number of hits in the ECal cluster to derive a calibration curve. We fit the distribution that is shown in Fig.~\ref{fig:07} with two second order polynomial functions, one for the region between 0\,-\,45 hits and the other for the remaining region of 45\,-\,200 hits. The choice of the fit functions is done based on practical reasons to have a reasonable match with the distribution. Then, we can correct for the bias using this fit. In Fig.~\ref{fig:08} we present the timing bias distribution for the $\tau_{\mathrm{fit}}$ estimator before and after calibration. We observe an improvement in the mean of the bias distribution.

In this paper we only familiarize the reader with the idea.
The full effect of a TOF calibration procedure on the mass reconstruction of charged hadrons is still to be investigated and will be addressed in further studies.
One needs to study and ensure that the calibration made for photons will hold for charged hadrons as their showers in the ECal have different properties which can impact the final TOF estimation. Also, the effects of the hit time resolution require a more detailed examination.

\section{Conclusions and outlook}

In this contribution, we presented how the reconstructed mass of charged hadrons depends on the choice of track parameters and the chosen TOF estimator. We conclude that using track parameters at the calorimeter surface $\Omega_{\mathrm{calo}}$, $\lambda_{\mathrm{calo}}$ and a TOF estimator based on fitting the time-of-arrival from hits in the first 10 layers of the Ecal shows the smallest mass bias, which also is of similar size for all studied charged hadrons. We present the idea to calibrate the TOF estimators based on photon clusters, for which the expected true time of arrival is simpler to predict than for hadrons. After a first attempt of such a calibration, a bias of about 1\,ps remains, which translates into a mass bias of about 3-4\,MeV for typical kaons. This is two orders of magnitude larger than the precision required to contribute to our knowledge of the kaon mass which is $O$(10\,keV).

In this study we focused on the assumption of a perfect time resolution.
However, the time resolution will affect the performance of each of the TOF estimators which may change the preferred one.
One needs to check the behavior of the TOF estimators with increasing finite time resolutions.
In addition, the implementation of a realistic digitizer is required to test TOF estimators in a more realistic environment, as the energy threshold of the electronics may cut off low-energy hit contributions which also impacts the performance.
As for now, taking the earliest MC contribution to the calorimeter hit does not show this effect.
An alternative option for the ECal TOF estimation would be to use the outermost Si tracker layer(s). This method has the advantage of being independent of the shower development and the corresponding material effects, but the disadvantage of only $O$(1) hit time measurement. This study has been performed with PFOs detected in the barrel section of the detector, however, a large fraction of PFOs going to the endcap region still remains unstudied due to the track length calculation which cannot account for multiple curls in one track. We plan to improve the track length calculation with the Kalman Filter to solve this problem and have more coverage region for PID.

\newpage
\section{Acknowledgements}
We would like to thank the LCC generator working group and the ILD software working group for providing the simulation and reconstruction tools and producing the Monte Carlo samples used in this study.
This work has benefited from computing services provided by the ILC Virtual Organization, supported by the national resource providers of the EGI Federation and the Open Science GRID. In this study we widely used the National Analysis Facility (NAF)~\cite{Haupt_2010}
and would like to thank Grid computational resources operated at Deutsches Elektronen-Synchrotron (DESY), Hamburg, Germany.
We thankfully acknowledge the support by the Deutsche Forschungsgemeinschaft (DFG, German Research Foundation) under Germany's Excellence Strategy EXC 2121 "Quantum Universe" 390833306.
\printbibliography[title=References]

\end{document}